\documentclass[10pt,twocolumn]{article}

\usepackage{isqcmc2025} 

\usepackage{graphicx,url}
\usepackage[braket, qm]{qcircuit}
\usepackage{color}

\usepackage[latin1]{inputenc}

\usepackage{amsmath}

\usepackage{anonymize} 
\usepackage{physics}
\usepackage{listings}

\lstdefinelanguage{OPENQASM}{
morekeywords={q, c}, 
morekeywords=[2]{OPENQASM, include}, 
emph={h,cx,qreg,creg,->}, 
morekeywords=[3]{measure}, 
morekeywords=[4]{barrier}, 
sensitive=true,
morecomment=[l]{//}, 
morestring=[b]",
literate={->}{{\textbf{\color{codeemph2}{$\to$}}}}1 
}

\title{Sonified Quantum Seizures. Sonification of time series in epileptic seizures and simulation of seizures via quantum modelling}

\author{\anonymize{Maria Mannone$^{\ast}$}\inst{1}\inst{2}\inst{3}\and
        \anonymize{Paulo Vitor Itabora\'i}\inst{4}\inst{5}
\and \anonymize{Omar Costa Hamido}\inst{6}
\and
\anonymize{Miriam Goldack} \inst{4}\inst{7}
\and
\anonymize{Norbert Marwan}\inst{2}\inst{3}\inst{8}\and\\
\anonymize{Peppino Fazio}\inst{9}\inst{{10}}
\and \anonymize{Patrizia Ribino}\inst{1}
}

\address{\anonymize{ICAR, National Research Council (CNR), Italy}
\nextinstitute
\anonymize{Institute of Physics and Astronomy, University of Potsdam, Germany}
\nextinstitute
\anonymize{Potsdam Institute for Climate Impact Research (PIK), Member of the Leibniz Association, Germany}
\nextinstitute
\anonymize{CQTA, Deutsches Elektronen-Synchrotron (DESY), Zeuthen, Germany}
\nextinstitute
\anonymize{CaSToRC, The Cyprus Institute, Nicosia, Cyprus}
\nextinstitute
\anonymize{Centre for Interdisciplinary Studies (CEIS20), University of Coimbra, Portugal}
\nextinstitute
\anonymize{Technical University of Berlin, Germany}
\nextinstitute
\anonymize{Institute of Geosciences, University of Potsdam, Germany}
\nextinstitute
\anonymize{DSMN, Ca' Foscari University of Venice, Italy} 
\nextinstitute
\anonymize{VSB - Technical University of Ostrava, Czechia}
\email{\anonymize{$^{\ast}$maria.mannone@uni-potsdam.de}}
}

\begin{document}

\maketitle

\begin{abstract}
We apply sonification strategies and quantum computing to the analysis of an episode of seizure. We first sonify the signal from a selection of channels (from real ECoG data), obtaining a polyphonic sequence. Then, we propose two quantum approaches to simulate a similar episode of seizure, and we sonify the results. The comparison of sonifications can give hints on similarities and discrepancies between real data and simulations, helping refine the \textit{in silico} model. This is a pioneering approach, showing how the combination of quantum computing and sonification can broaden the perspective of real-data investigation, and helping define \textcolor{black}{a} new test bench for analysis and prediction of seizures.
\end{abstract}

\section{Introduction}\label{introduction}

Prediction is crucial for epilepsy \cite{lekshmy2022comparative}, and it is often based on extensive data from electroencephalography (EEG) 
 \cite{maimaiti2022overview}. Typology of EEGs can be distinguished \cite{wang2024research} into time-domain \cite{direito2017realistic,sun2022continuous}, frequency-domain  \cite{savadkoohi2020machine,wang2020expert}, and time-frequency \cite{yu2020epileptic,le2017surface} analysis.
Some invasive EEG (iEEG) can also be performed in preparation \textcolor{black}{for} surgery; it is the case of electrocorticography (ECoG), used to identify the precise source of seizure, before proceeding with surgical resection of particular connections \cite{balaji}.
It is also known that the degree of synchronisation presents variations across various brain regions
\cite{varotto2012epileptogenic,tomlinson2017interictal}.

Given the ``choral'' character of brain signal, the simultaneous signal from different areas, and the effect of propagation of amplitude-frequency oscillation during a seizure through different parts of the brain, we can \textcolor{black}{consider} an auditory equivalent of the process. Also exploiting our ability to analyze \textcolor{black}{the} complex structure of music, we can use a sonification strategy to obtain an auditory representation of \textcolor{black}{a} multi-channel signal from EEG.
\textcolor{black}{Moreover}, given the probabilistic nature of seizures, we propose here to model a toy application, starting from the example of real data, using the intrinsic probability of quantum circuits. We could simulate the pre-, during-, and post-seizure sequence of a few channels, considering the onset, development, and end of a seizure as an instance of a wavelet. 
\textcolor{black}{In particular, we introduce two quantum-based methods to simulate seizure-like events and apply sonification to the results. By comparing these sonifications with real-world data, we can identify similarities and differences that aid in refining the \textit{in silico} model. This innovative approach demonstrates how merging quantum computing with sonification can expand the scope of real-data analysis, establishing a novel framework for seizure prediction and evaluation.}

\section{A case study of time series and the sonification}\label{time_series_sonification}

\begin{figure}[ht!]
\centering
\includegraphics[width=\columnwidth]{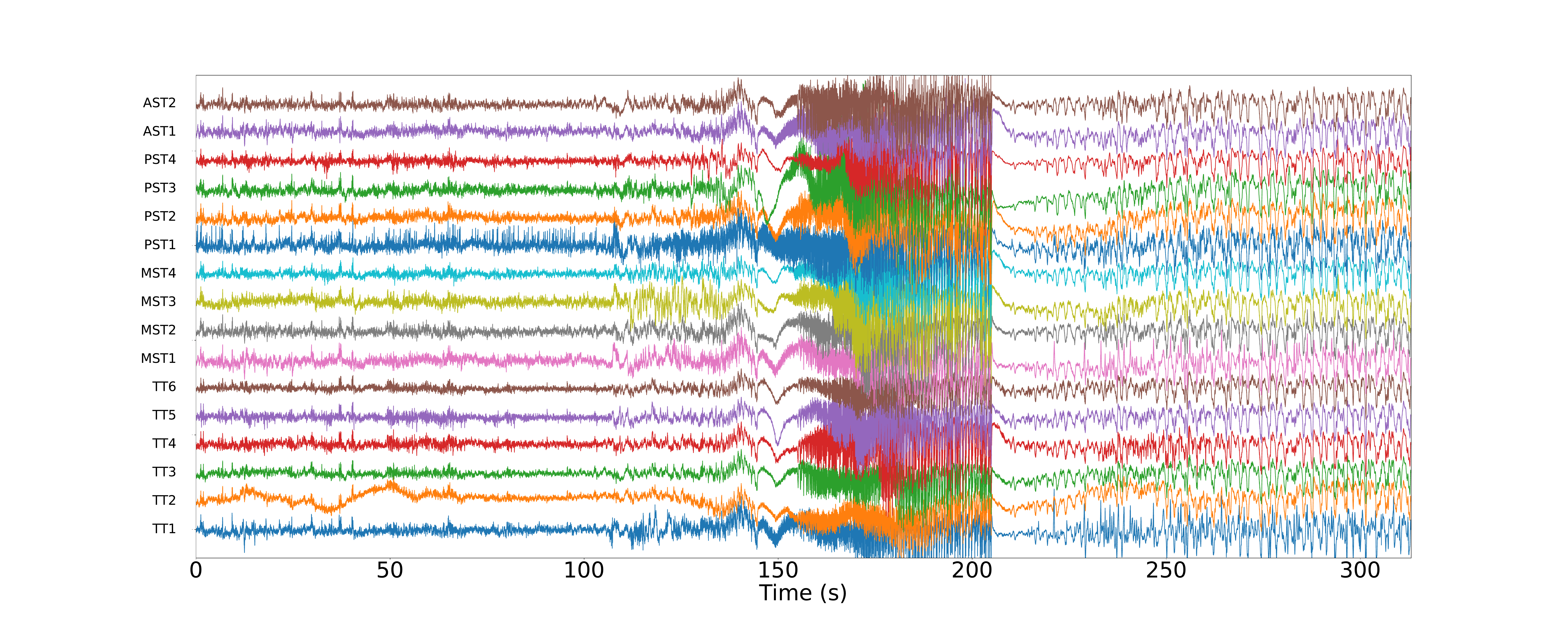}
\caption{Signal from a selection of channels for pre-, during, and post-epileptic seizure, for a real case study.}\label{iEEG_data_selected_few_no_annotations}
\end{figure}

We consider here the time series for a real case of pre-, during, and post-epileptic seizure of a patient. We present the choice of patient and the selection of channels according to the article \cite{epilepsy_ASPAI}. The case study is an electro-corticography (a pre-surgical invasive analysis, ECoG) from the dataset \textit{Fragility Multi-Center Retrospective Study}. The patient (ID 02) is a woman of 28 years old, with a hypothesised left anterior temporal lobe epilepsy. We focus on the electrodes corresponding to the spreading of the seizure, also following the annotations within the dataset. The time is measured in seconds. The seizure shows an early onset at 105.90s, and it is mostly detected on the TT1 channel, and then on \textcolor{black}{the} PST. After the slowing of the rhythm at 109.78s (channel G1-3), at 154.97, there is a seizure spread involving TT, AST, MST, and PST electrodes that starts. A more general spread starts at 164.85s. Finally, the offset of the seizure is indicated at 204.74s.
Figure \ref{iEEG_data_selected_few_no_annotations} shows the signal from the selected channels.

\section{Sonification of the epilepsy data}

To obtain a simple yet effective sonification, we map time series to pitch-onset series. For reasons of computational limits, from the original time series, a sampling with only one value each 500 is considered and mapped to sound. We consider simple sinusoidal functions. The duration of each note was set equal to $0.1$ seconds, shortening the overall sequence time. The $y$-values are mapped to pitches between min\_freq  
of $261.63$ \textcolor{black}{and} max\_freq $1244.51$ Hz, corresponding to the frequency range between C4 and B5.

\begin{figure}[ht!]
\centering
\includegraphics[width=\columnwidth]{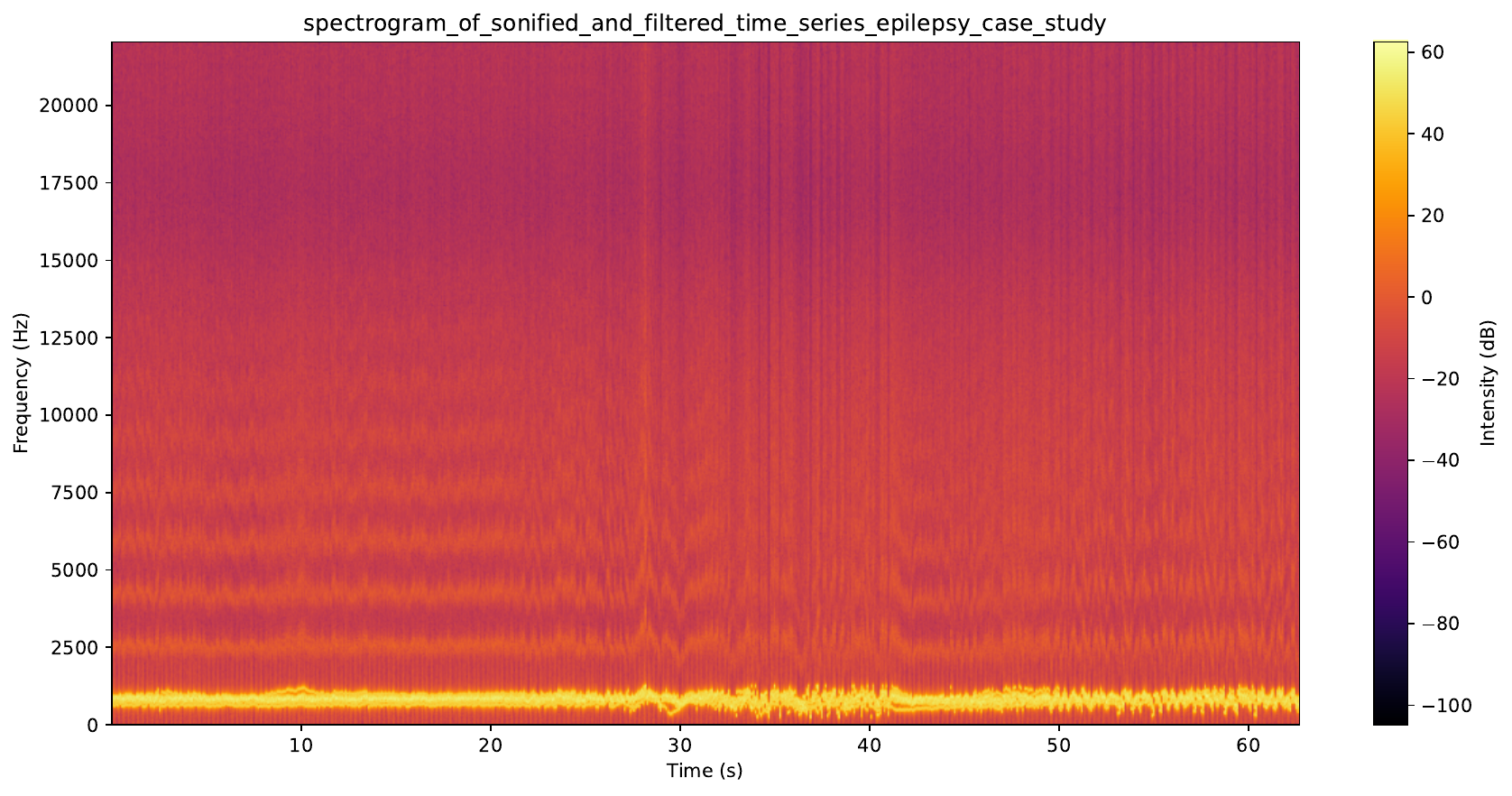}
\caption{Sonified signal from the channels of Figure \ref{iEEG_data_selected_few_no_annotations}.}\label{spectrogram_of_sonified_and_filtered_time_series_epilepsy_case_study}
\end{figure}

We thus adopt the simplest sonification strategy possible, \textcolor{black}{which}, however, already creates, in the mind of the listener, a feeling of alteration during seizure.
The sonification is performed in Python (Jupyter Notebook). A low-pass filter and de-noise steps helped clean the sound. We obtain a polyphonic sound sequence, with one voice for each channel, with spectrogram shown in Figure \ref{spectrogram_of_sonified_and_filtered_time_series_epilepsy_case_study}. An alteration of the pattern signals the beginning and ending of the seizure; it is audible between 120s and 200s approximately, and also visible in the spectrogram; similarly for the post-seizure high-amplitude pattern.

\section{Enhancing the Sonification by Integrating quantum algorithms}\label{quantum_application}

To integrate quantum computing within a musical \textit{praxis}, we investigate two separate quantum computer music techniques applied to the current case study, augmenting the dataset to enrich the layers of sonification with correlated gestures.

\subsection{Technique 1: Quantum Information Retrieval}

The first approach draws inspiration from Quantum Signal Processing techniques, with the end of computing information metrics regarding a certain signal. 

For quantum circuits, a common way of extracting metrics from a statevector is by estimating the expectation values of specific quantum observables.

In this case, we encoded the MST4 channel of the ECoG data as a signal to compute the expectation value of statistical moments of a sliding window block onto the signal. 
More specifically, we have used a form of amplitude encoding, that is, Quantum Probability Amplitude Modulation (QPAM in the quantum music literature) to compute the moving 4th statistical momentum, Eq. \eqref{eq:rollingkurtosis2}, of a 16-sampled window with hop size 10, see Eq. \eqref{eq:rollingkurtosis}.
\begin{align}
\label{eq:rollingkurtosis0}
\ket{\psi_n} = \sum_{i=0}^{15} a_{n+i}\ket{i}\; &; \;\; n \in \{0, 10, 20, ...,  N-16\}.
\\
    \label{eq:rollingkurtosis}
    \hat{M}_4 &= \sum_{i=0}^{16} i^4 \ket{i}\bra{i} 
\\
    \label{eq:rollingkurtosis2}
    \langle \hat{M_4}\rangle_n &= \sum_{i=0}^{15} i^4 \abs{a_{n+i}}^2     
\end{align}

The in-depth methodology of computing the statistical moments of QPAM signals, as well as the observable and quantum circuit design, is discussed \textcolor{black}{in} detail in an unpublished work by \textcolor{red}{[Manuscript in Preparation].} This method is applied here to sonify the kurtosis of the MST4 channel. 
This circuit and the estimation were simulated using qiskit\footnote{\url{https://www.ibm.com/quantum/qiskit}} with a Statevector simulator. 

\begin{figure}[h]
\centering
  \includegraphics[width=.45\textwidth]{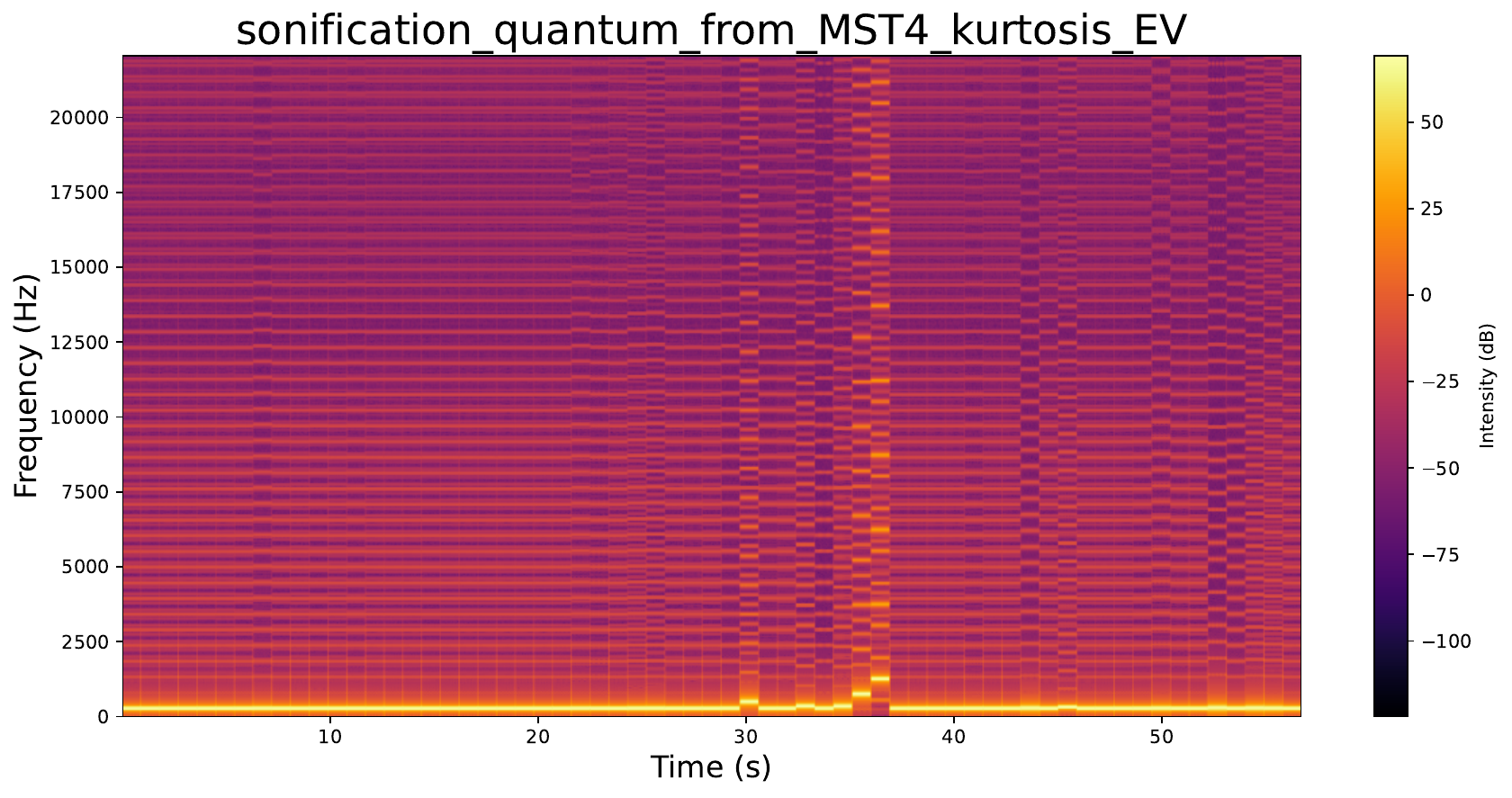}
\caption{Spectrogram of the sonified Rolling expectation value of the kurtosis of the MST4 channel encoded using a QPAM approach, simulated.}
\label{fig:kurtosis_q}
\end{figure}

The result depicted in Figure \ref{fig:kurtosis_q} shows an example of the kurtosis mapped to the frequency of an oscillator. It can be noticed how the 4th momentum retains information about large oscillations of the MST4 signal, capturing the onset of the crisis, characterizing its end, as well as indicating a remaining instability of the channel after the episode. As a result, this signal can be mapped to other musical parameters, as an envelope that retains the global characteristics of the seizure episode. It can complement the sonification in Figure \ref{time_series_sonification}. Comparing the spectrogram of this reconstructed channel with the spectrogram of the sonification of the original ECoG sequences, slightly longer (Figure \ref{spectrogram_of_sonified_and_filtered_time_series_epilepsy_case_study}), we notice the correspondence of seizure onset slightly before 30s, a second peak before 40s, and, after the seizure (after 40s), a general higher activity within the channel.

\subsection{Technique 2: Quantum Modelling and Emulation}

The second approach comes from a substantially different route. Instead of attempting to encode and process the real ECoG data in a Quantum Signal Processing fashion, in this case, we utilize the dataset to drive the design of a quantum system that can be simulated through a quantum circuit. The designed circuit may also display complex and inter-connected behaviour, mirroring (emulating) some patterns seen in ECoG.

For an artistic application, this technique could be interesting for the creation of musical counterpoints. In other words, we build a new dataset that ``reacts'' to the changes of the original one.

For this technique, we explore the Hamiltonian in Eq. \eqref{eq:ising}, describing a time-dependent, 1D-Transverse Field Ising model with nearest-neighbour interaction.
\begin{equation}\label{eq:ising}
    H(t) = \sum_{n=0}^{N-1}J_{n}(t)\, \sigma^z_n\sigma^z_{n+1} + \sum_{n=0}^{N-1}h_x(t) \sigma^x_n
\end{equation}

If we define a chain of 16 spins, we can then encode 16 channels of the EcoG dataset to express the coupling coefficients $J_n$ over time. Additionally, we exploit the result of Technique 1 and map the 4th momentum of the MST4 channel to change the transverse field $h_x$.

With Eq. \eqref{eq:ising}, we can perform and study a trotterised \cite{suzuki1976generalized,trotter1959product} time evolution operation on the system, given an initial condition. The approximation is given in equation \eqref{eq:trotter_product2}. This approach has been explored in the visual arts within the QC-PAINT Quantum Transformations painting project. The reader can find a detailed explanation of the trotterisation process and the quantum circuit design in their work \cite{crippa2025quantumcomputinginspiredpaintings}. 

The small difference in the context of this work (compared to QC-PAINT) is that the Hamiltonian is varying at each trotter step. This was approximated  \footnote{Please note that varying the Hamiltonian within a trotter step may harm the trotter product approximation and hence the physical interpretation of an Ising Model. For a proper physical interpretation, one needs to ensure that the variation of the time-dependent Hamiltonian is also bounded by a small variation, as described in the Adaptative Troterisation method \cite{Zhao2024}. However, this was not taken into account in this work.} using the expressions in Eq. \eqref{eq:trotter_product2}
 in Appendix \ref{sec:appendix}, leading to Eq. \eqref{eq:totter_ising}:
\begin{align}\label{eq:totter_ising}
    \ket{\psi(t)} = U(t)\ket{\psi(0)} \approx \prod^k \prod_{n=0}^{N-1} e^{-iH_n(k) t/k}.
\end{align}

\begin{figure}[h]
\centering
\scalebox{0.7}{
\Qcircuit @C=0.7em @R=0.1em @!R { \\
	 	\nghost{{q}_{0} :  } & \lstick{{q}_{0} :  } & \ctrl{1} & \qw & \ctrl{1} & \qw & \targ & \gate{\mathrm{R_Z}\,(\mathrm{J_{15}})} & \qw & \targ  & \gate{\mathrm{R_X}\,(\mathrm{h_x})} & \qw & \qw\\
	 	\nghost{{q}_{1} :  } & \lstick{{q}_{1} :  } & \targ & \gate{\mathrm{R_Z}\,(\mathrm{J_0})} & \targ & \ctrl{1} & \qw & \qw & \ctrl{1} & \qw  & \gate{\mathrm{R_X}\,(\mathrm{h_x})} & \qw & \qw\\
	 	\nghost{{q}_{2} :  } & \lstick{{q}_{2} :  } & \ctrl{1} & \qw & \ctrl{1} & \targ & \qw & \gate{\mathrm{R_Z}\,(\mathrm{J_1})} & \targ & \qw & \gate{\mathrm{R_X}\,(\mathrm{h_x})} & \qw & \qw\\
	 	\nghost{{q}_{3} :  } & \lstick{{q}_{3} :  } & \targ & \gate{\mathrm{R_Z}\,(\mathrm{J_2})} & \targ & \ctrl{1} & \qw & \qw & \ctrl{1} & \qw & \gate{\mathrm{R_X}\,(\mathrm{h_x})} & \qw & \qw\\
	 	\nghost{{q}_{4} :  } & \lstick{{q}_{4} :  } & \ctrl{1} & \qw & \ctrl{1} & \targ & \qw & \gate{\mathrm{R_Z}\,(\mathrm{J_3})} & \targ & \qw & \gate{\mathrm{R_X}\,(\mathrm{h_x})} & \qw & \qw\\
	 	\nghost{{q}_{5} :  } & \lstick{{q}_{5} :  } & \targ & \gate{\mathrm{R_Z}\,(\mathrm{J_4})} & \targ & \ctrl{1} & \qw & \qw & \ctrl{1} & \qw & \gate{\mathrm{R_X}\,(\mathrm{h_x})} & \qw & \qw\\
	 	\nghost{{q}_{6} :  } & \lstick{{q}_{6} :  } & \ctrl{1} & \qw & \ctrl{1} & \targ & \qw & \gate{\mathrm{R_Z}\,(\mathrm{J_5})} & \targ & \qw  & \gate{\mathrm{R_X}\,(\mathrm{h_x})} & \qw & \qw\\
	 	\nghost{{q}_{7} :  } & \lstick{{q}_{7} :  } & \targ & \gate{\mathrm{R_Z}\,(\mathrm{J_6})} & \targ & \ctrl{1} & \qw & \qw & \ctrl{1} & \qw & \gate{\mathrm{R_X}\,(\mathrm{h_x})} & \qw & \qw\\
	 	\nghost{{q}_{8} :  } & \lstick{{q}_{8} :  } & \ctrl{1} & \qw & \ctrl{1} & \targ & \qw & \gate{\mathrm{R_Z}\,(\mathrm{J_7})} & \targ & \qw & \gate{\mathrm{R_X}\,(\mathrm{h_x})} & \qw & \qw\\
	 	\nghost{{q}_{9} :  } & \lstick{{q}_{9} :  } & \targ & \gate{\mathrm{R_Z}\,(\mathrm{J_8})} & \targ & \ctrl{1} & \qw & \qw & \ctrl{1} & \qw & \gate{\mathrm{R_X}\,(\mathrm{h_x})} & \qw & \qw\\
	 	\nghost{{q}_{10} :  } & \lstick{{q}_{10} :  } & \ctrl{1} & \qw & \ctrl{1} & \targ & \qw & \gate{\mathrm{R_Z}\,(\mathrm{J_9})} & \targ & \qw & \gate{\mathrm{R_X}\,(\mathrm{h_x})} & \qw & \qw\\
	 	\nghost{{q}_{11} :  } & \lstick{{q}_{11} :  } & \targ & \gate{\mathrm{R_Z}\,(\mathrm{J_{10}})} & \targ & \ctrl{1} & \qw & \qw & \ctrl{1} & \qw & \gate{\mathrm{R_X}\,(\mathrm{h_x})} & \qw & \qw\\
	 	\nghost{{q}_{12} :  } & \lstick{{q}_{12} :  } & \ctrl{1} & \qw & \ctrl{1} & \targ & \qw & \gate{\mathrm{R_Z}\,(\mathrm{J_{11}})} & \targ & \qw & \gate{\mathrm{R_X}\,(\mathrm{h_x})} & \qw & \qw\\
	 	\nghost{{q}_{13} :  } & \lstick{{q}_{13} :  } & \targ & \gate{\mathrm{R_Z}\,(\mathrm{J_{12}})} & \targ & \ctrl{1} & \qw & \qw & \ctrl{1} & \qw & \gate{\mathrm{R_X}\,(\mathrm{h_x})} & \qw & \qw\\
	 	\nghost{{q}_{14} :  } & \lstick{{q}_{14} :  } & \ctrl{1} & \qw & \ctrl{1} & \targ & \qw & \gate{\mathrm{R_Z}\,(\mathrm{J_{13}})} & \targ & \qw & \gate{\mathrm{R_X}\,(\mathrm{h_x})} & \qw & \qw\\
	 	\nghost{{q}_{15} :  } & \lstick{{q}_{15} :  } & \targ & \gate{\mathrm{R_Z}\,(\mathrm{J_{14}})} & \targ & \qw & \ctrl{-15} & \qw & \qw & \ctrl{-15} & \gate{\mathrm{R_X}\,(\mathrm{h_x})} & \qw & \qw\\
\\ }}
\caption{Circuit of a single time evolution step.}\label{fig:trotter_circuit}
\end{figure}
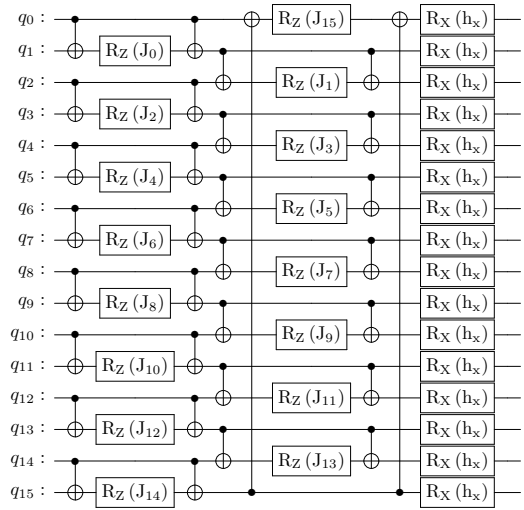

For example, the circuit to obtain the time evolution for the first time step is shown in Figure \ref{fig:trotter_circuit}.

Finally, we simulate the circuit to obtain an estimation of the expectation value of the observable defined in Eq. \eqref{eq:observable}, for each qubit $n$.
\begin{equation}\label{eq:observable}
    \hat{O}_n = \frac{(I-\sigma_n^z)}{2}     \;\; ;\;\; \hat{O}_n\ket{0}=0 \ \ \ \text{and} \ \ \ \hat{O}_n \ket{1}=\ket{1}
\end{equation}

Notice that the expectation value of the described observable $\langle{\hat{O}_n}\rangle$ could be interpreted as the probability of measuring qubit $n$ in the $\ket{1}$ state. For instance, we can consider an arbitrary quantum state, as in Eq. \eqref{eq:arbitrarystate}.
\begin{equation}\label{eq:arbitrarystate}
    \ket{\psi} = \alpha \ket0 + \beta \ket1
\end{equation}
The expectation value would then be described by Eq. \eqref{eq:expval}, where $\bar\alpha,\,\bar\beta$ indicate the complex conjugate of $\alpha,\,\beta$, respectively.
\begin{equation}
\begin{aligned}\label{eq:expval}
    \bra{\psi}\hat{O}_n\ket{\psi} = \abs{\alpha}^2\bra{0}\hat{O}_n\ket{0} + \abs{\beta}^2\bra{1}\hat{O}_n\ket{1}+\\
    \alpha\bar\beta\bra{1}\hat{O}_n\ket{0} + \bar\alpha\beta\bra{0}\hat{O}_n\ket{1}+\\
    = \abs{\beta}^2\braket{1}{1} = P(n=\ket{1})
\end{aligned}
\end{equation}
By estimating the referred quantity for each qubit, we are effectively computing the \textit{marginal distribution} of the quantum state. 

\begin{figure}[h]
\centering
  \includegraphics[width=.45\textwidth]{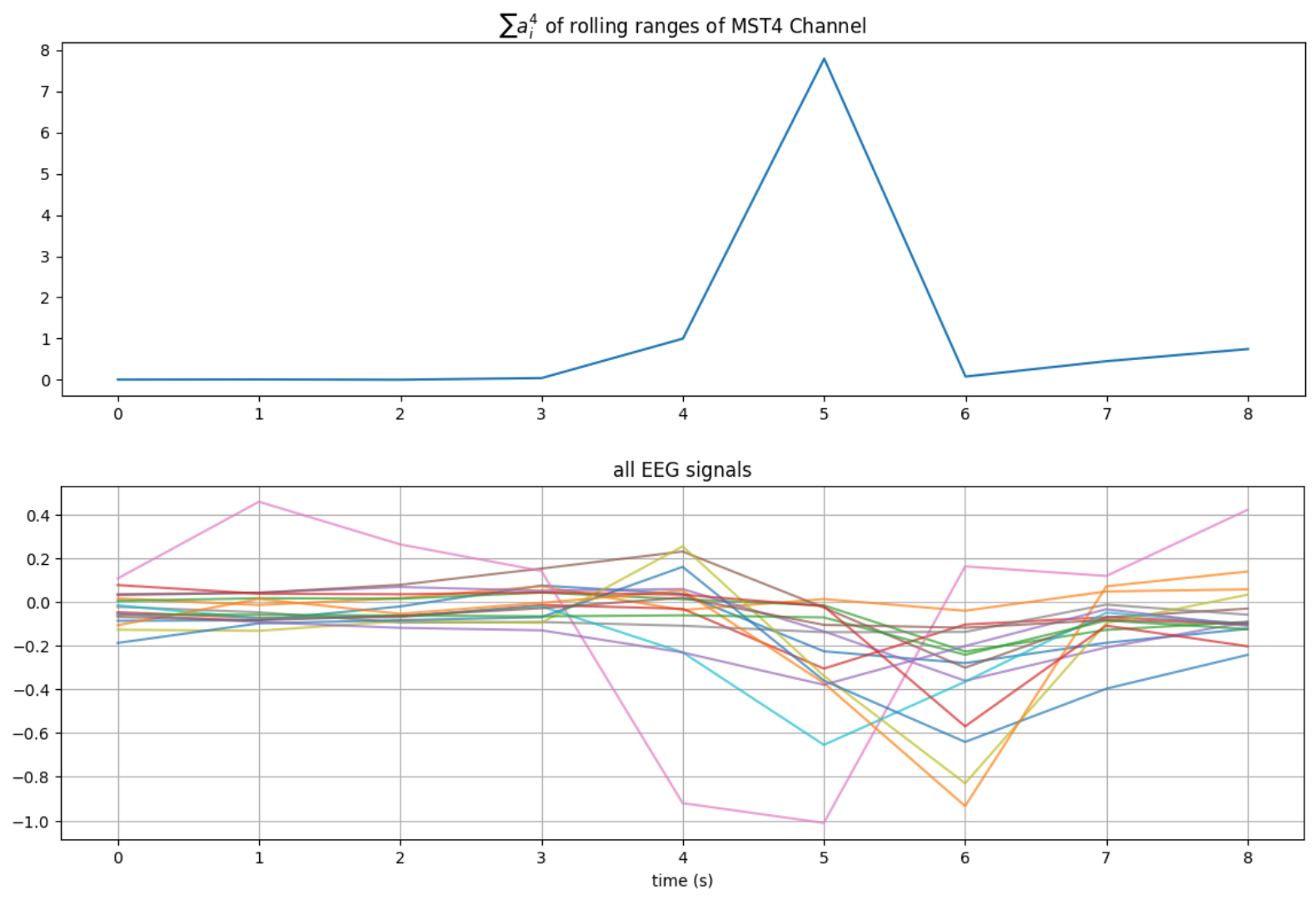}
\caption{Reduced, averaged and renormalised dataset for 9 time steps. Top: The outcome of technique 1 ($\langle a^4\rangle$ for MST4). At $t=4$, the amplitude is designed to be exactly 1 (phase transition). Bottom: 16 channels of the ECoG dataset from Figure \ref{time_series_sonification}.}
\label{fig:data_redux}
\end{figure}

Nevertheless, performing a large time evolution with many time steps requires the inclusion of one circuit layer per time step, which would be unfeasible for $\approx10^5$ time steps. Hence, as a proof of concept, we reduce the datasets by accumulating and averaging the time series into 9 evenly-spaced segments (Figure \ref{fig:data_redux}). 

Furthermore, the signals are renormalised so that the coupling coefficients are weakly bounded to a $[-1, 1]$ range in the beginning, allowing for larger variations to appear when the seizure episode starts. As for the transverse-field signal, it was renormalised such that a phase transition  ($h_x =1$) would occur exactly at the fourth timestep (Figure \ref{fig:data_redux}, top).

From Eq. \eqref{eq:totter_ising}, we then take a time-step length of ${t}/{k}=0.5$ and $k=9$ trotter steps. Additionally, we use $N=16$ qubits for encoding 16 channels. The proof-of-concept simulation was run using a Qiskit-Aer estimator, for the observables in Eq. \eqref{eq:observable}, computed for each qubit.

\begin{figure}[h]
\centering
  \includegraphics[width=.45\textwidth]{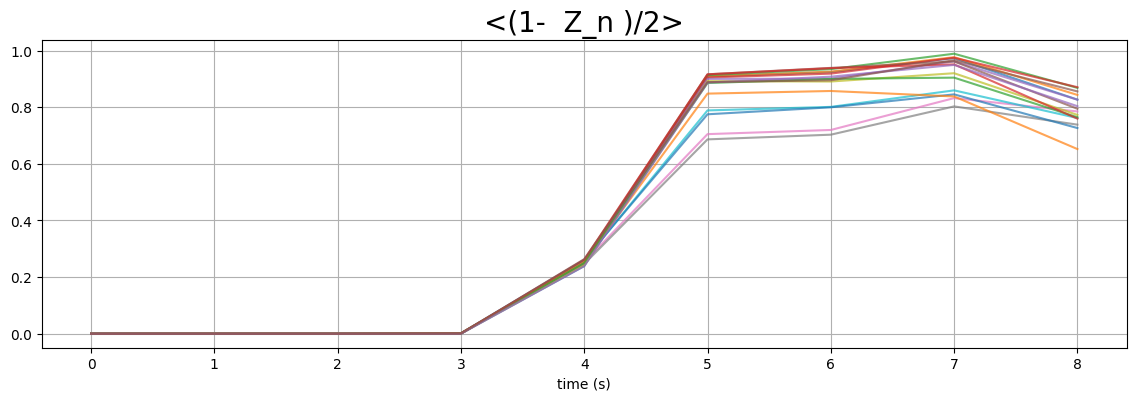}
\caption{Marginal distribution of the time evolution of the hamiltonian in Eq. \eqref{eq:ising}}
\label{fig:isingevo}
\end{figure}

The results are depicted in Figure \ref{fig:isingevo}, where a clear phase transition can be observed, as expected. 

\begin{figure}[h]
\centering
  \includegraphics[width=.45\textwidth]{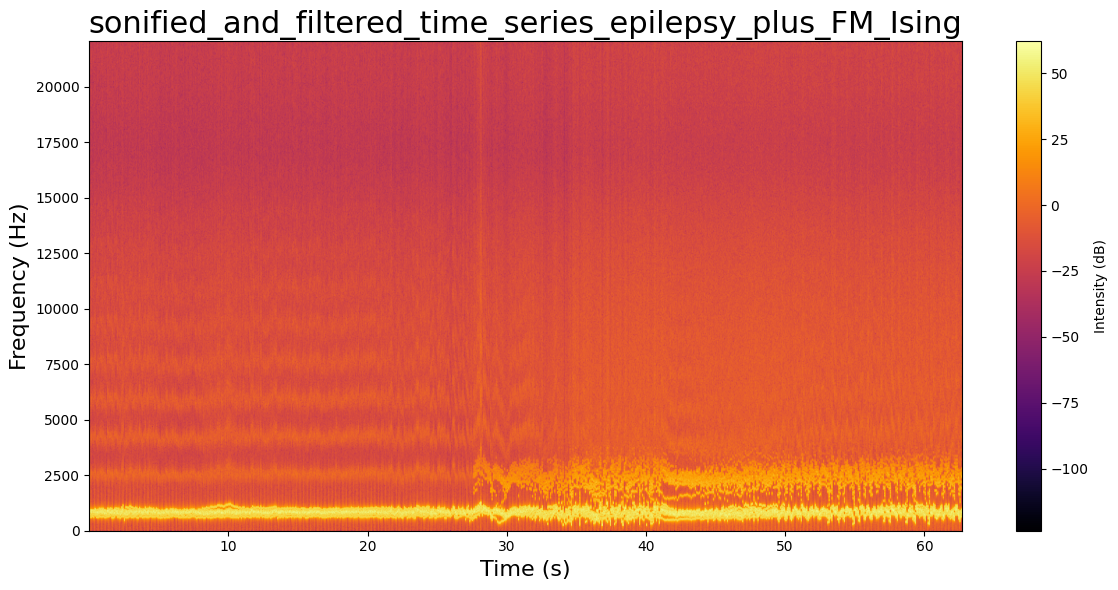}
\caption{Sonified ECoG signals, modulated by the Ising time evolution}
\label{fig:Sonified_signal_plus_FM}
\end{figure}

Finally, to enhance the sonification in Figure \ref{spectrogram_of_sonified_and_filtered_time_series_epilepsy_case_study}, we have introduced a Frequency Modulation (FM) into the original mapping. In turn, the Ising phase transition controlled the modulation index of the FM, leading to the result depicted in Figure \ref{fig:Sonified_signal_plus_FM}, where the generated data enhances the identification of the seizure onset, while also indicating the continued increased activity post-crisis.

\section{Discussions and conclusions}\label{discussions}

In this article, we first proposed a sonification of real-data epilepsy from a human patient. Then, we exploited data to develop a quantum-based modelling, considering different strategies, including the Ising time-evolution model. Finally, we sonified the different outputs and \textcolor{black}{discussed} them.

More refined sonifications can include variations of timbre and loudness, to produce illusions of different directionality, to really create in the mind of the listener the feeling of the abnormal activation of different brain areas during a seizure. This would constitute a fruitful application of the conjecture of gestural similarity, seeing music as the result of a movement in space and time, and establishing perceptive analogies between different objects and their transformations through sound \cite{gest_sim}.

Although computing statistical metrics using a quantum circuit may prove to be rather inefficient for the considered case and scale, as one can easily compute statistical moments with a classical machine for a large set of objects. However, such a first attempt within a creative framework could provide fertile ground for exploring new potentials and extensions of such algorithms, while allowing a larger community to learn and interact with quantum computing.

In the Ising time-evolution result, it can be noted that there were insufficient time steps or resolution for the Ising lattice to allow a transition to the post-crisis state. Further work could explore finer resolutions for the time steps.

Future research will focus on tuning the quantum approach to develop a predictive system.
In fact, quantum computing can also be used to model classical phenomena, using probability amplitudes and suitably defining states.

\section{Code availability}

The code to read ECoG signal, written for the study in \cite{epilepsy_ASPAI}, can be accessed here: \url{https://github.com/medusamedusa/K-operator_epilepsy}. The code to sonify the files is available at \url{ https://github.com/medusamedusa/sonification_epilepsy}

\section{Data availability}

Data concern patient 02 from the dataset \textit{Fragility Multi-Center Retrospective Study}, and they can be accessed from the OpenNeuro platform at:
\url{https://openneuro.org/datasets/ds003029/versions/1.0.6}.

\section{Acknowledgment}

The work of \anonymize{M.M. and P.R.} is supported by the project funded by \anonymize{Next Generation EU -- ``Age-It -- Ageing well in an ageing society'' project (PE0000015), National Recovery and Resilience Plan (NRRP) -- PE8 -- Mission 4, C2, Intervention 1.3, CUP B83C22004880006}.

The research by \anonymize{P.F.} was supported by the \anonymize{European Union within the REFRESH project -- Research} 
\anonymize{Excellence For Region Sustainability and High-tech Industries ID No. CZ.10.03.01/00/22 003/0000048 of the European Just Transition Fund}.

\anonymize{O.C.H.} is supported by \anonymize{the European Project IIMPAQCT, under grant agreement Nr. 101109258 (DOI \texttt{10.3030/101109258})}.

The work of \anonymize{P.I.} is funded by \anonymize{the European Union's Horizon Europe Framework Programme (HORIZON) under the ERA Chair scheme with grant agreement no. 101087126}.

\anonymize{P.I. and M.G.} are supported with funds from \anonymize{the Ministry of Science, Research and Culture of the State of Brandenburg within the Centre for Quantum Technologies and Applications (CQTA)}.

\begin{center}
    \anonimg[width = 0.08\textwidth]{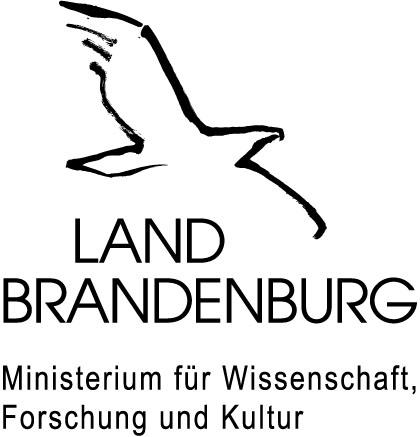}
\end{center}

\bibliographystyle{unsrt}
\bibliography{example}

\appendix
\section{Appendix: Trotterisation}
\label{sec:appendix}
The Trotter product considers an approximation of matrix exponents where quadratic terms are considered sufficiently small. The approximations below, in Eq. \eqref{eq:trotter_product2}, are discussed and detailed in \cite{kluber2023trotterization}.

\begin{equation}\label{eq:trotter_product2}
\begin{split}
&e^{A+B} \approx\left( e^{A / N} e^{B / N} \right)^{N} \\ 
&\left( e^{A / N} e^{B / N} \right)^{N}\approx\left[ \left( I+\frac{A} {N} \right)\! \left( I+\frac{B} {N} \right) \right]^{N} \\ 
&\left[\! \left( I+\frac{A} {N} \right)\!\! \left( I+\frac{B} {N} \right)\! \right]^{N}\!\!\!\!\!=\left[ I+\frac{A} {N}+\frac{B}{N}+\frac{A B} {N^{2}} \right]^{N} \\ 
&  \left[ I+\frac{A} {N}+\frac{B} {N}+\frac{A B} {N^{2}} \right]^{N}\approx\left[ I+\frac{A+B} {N} \right]^{N}
\end{split}   
\end{equation}

\end{document}